\documentclass[final,3p,times,longtitle]{elsarticle}

\usepackage{ecrc}

\volume{00}

\firstpage{1}

\journalname{Physics Procedia}

\runauth{The {\sc Majorana} Collaboration}

\jid{procs}

\jnltitlelogo{Physics Procedia}

\CopyrightLine{2011}{Published by Elsevier Ltd.}

\usepackage{amssymb}
\begin{document}

% Definitions...
\def\nuc#1#2{${}^{#1}$#2}
\def\BBz{0$\nu\beta\beta$}
\def\BBt{2$\nu\beta\beta$}
\def\BB{$\beta\beta$}
\def\Tz{$T^{0\nu}_{1/2}$}
\def\Tt{$T^{2\nu}_{1/2}$}
\def\mj{M{\sc ajo\-ra\-na}}
\def\dem{D{\sc e\-mon\-strat\-or}}
\def\mg{M{\sc a}G{\sc e}}
\def\QBB{Q$_{\beta\beta}$}
\def\mBB{$\left < \mbox{m}_{\beta\beta} \right >$}
\def\ge{$^{76}$Ge}

\begin{frontmatter}

\title{Background Model for the \textsc{Majorana Demonstrator}}

\author[uw]{C. Cuesta}	

\author[lbnl]{N.~Abgrall}		
\author[pnnl]{E.~Aguayo}
\author[usc,ornl]{F.T.~Avignone~III}
\author[ITEP]{A.S.~Barabash}	
\author[ornl]{F.E.~Bertrand}
\author[lanl]{M.~Boswell}
\author[JINR]{V.~Brudanin}
\author[duke,tunl]{M.~Busch}	
\author[usd]{D.~Byram}
\author[sdsmt]{A.S.~Caldwell}
\author[lbnl]{Y-D.~Chan}
\author[sdsmt]{C.D.~Christofferson}
\author[ncsu,tunl]{D.C.~Combs}

\author[uw]{J.A.~Detwiler}	
\author[uw]{P.J.~Doe}
\author[ut]{Yu.~Efremenko}
\author[JINR]{V.~Egorov}
\author[ou]{H.~Ejiri}
\author[lanl]{S.R.~Elliott}
\author[pnnl]{J.E.~Fast}
\author[unc,tunl]{P.~Finnerty}
\author[unc,tunl]{F.M.~Fraenkle}
\author[ornl]{A.~Galindo-Uribarri}	
\author[unc,tunl]{G.K.~Giovanetti}
\author[lanl]{J. Goett}	
\author[ornl]{M.P.~Green}
\author[uw]{J. Gruszko}		
\author[usc]{V.E.~Guiseppe}	
\author[JINR]{K.~Gusev}
\author[alberta]{A.L.~Hallin}
\author[ou]{R.~Hazama}
\author[lbnl]{A.~Hegai\fnref{TU}}
\author[unc,tunl]{R.~Henning}
\author[pnnl]{E.W.~Hoppe}
\author[sdsmt]{S. Howard}
\author[unc,tunl]{M.A.~Howe}
\author[blhill]{K.J.~Keeter}
\author[ttu]{M.F.~Kidd}	
\author[JINR]{O.~Kochetov}
\author[ITEP]{S.I.~Konovalov}
\author[pnnl]{R.T.~Kouzes}
\author[pnnl]{B.D.~LaFerriere}
\author[uw]{J.~Leon}	
\author[ncsu,tunl]{L.E.~Leviner}
\author[sjtu]{J.C.~Loach}	
\author[unc,tunl]{J.~MacMullin}
\author[unc,tunl]{S.~MacMullin}
\author[usd]{R.D.~Martin}
\author[unc,tunl]{S. Meijer}	
\author[lbnl]{S.~Mertens}		
\author[ou]{M.~Nomachi}
\author[pnnl]{J.L.~Orrell}
\author[unc,tunl]{C. O'Shaughnessy}	
\author[pnnl]{N.R.~Overman}
\author[ncsu,tunl]{D.G.~Phillips~II}
\author[lbnl]{A.W.P.~Poon}
\author[usd]{K.~Pushkin}
\author[ornl]{D.C.~Radford}
\author[unc,tunl]{J.~Rager}	
\author[lanl]{K.~Rielage}
\author[uw]{R.G.H.~Robertson}
\author[ut,ornl]{E.~Romero-Romero}
\author[lanl]{M.C.~Ronquest}	
\author[uw]{A.G.~Schubert}		
\author[unc,tunl]{B.~Shanks}	
\author[ou]{T.~Shima}
\author[JINR]{M.~Shirchenko}
\author[unc,tunl]{K.J.~Snavely}	
\author[usd]{N.~Snyder}	
\author[sdsmt]{A.M.~Suriano}
\author[blhill,sdsmt]{J.~Thompson}
\author[JINR]{V.~Timkin}
\author[duke,tunl]{W.~Tornow}
\author[unc,tunl]{J.E.~Trimble}
\author[ornl]{R.L.~Varner}
\author[ut]{S.~Vasilyev}
\author[lbnl]{K.~Vetter\fnref{ucb}}
\author[unc,tunl]{K.~Vorren}
\author[ornl]{B.R.~White}	
\author[unc,tunl,ornl]{J.F.~Wilkerson}
\author[usc]{C.~Wiseman}		
\author[lanl]{W.~Xu}
\author[JINR]{E.~Yakushev}
\author[ncsu,tunl]{A.R.~Young}
\author[ornl]{C.-H.~Yu}
\author[ITEP]{V.~Yumatov}

\address[uw]{Center for Experimental Nuclear Physics and Astrophysics, and Department of Physics, University of Washington, Seattle, WA, USA}

\address[lbnl]{Nuclear Science Division, Lawrence Berkeley National Laboratory, Berkeley, CA, USA}
\address[pnnl]{Pacific Northwest National Laboratory, Richland, WA, USA}
\address[usc]{Department of Physics and Astronomy, University of South Carolina, Columbia, SC, USA}
\address[ornl]{Oak Ridge National Laboratory, Oak Ridge, TN, USA}
\address[ITEP]{Institute for Theoretical and Experimental Physics, Moscow, Russia}
\address[lanl]{Los Alamos National Laboratory, Los Alamos, NM, USA}
\address[JINR]{Joint Institute for Nuclear Research, Dubna, Russia}
\address[duke]{Department of Physics, Duke University, Durham, NC, USA}
\address[tunl]{Triangle Universities Nuclear Laboratory, Durham, NC, USA}
\address[usd]{Department of Physics, University of South Dakota, Vermillion, SD, USA}
\address[sdsmt]{South Dakota School of Mines and Technology, Rapid City, SD, USA}
\address[ncsu]{Department of Physics, North Carolina State University, Raleigh, NC, USA}

\address[ut]{Department of Physics and Astronomy, University of Tennessee, Knoxville, TN, USA}
\address[ou]{Research Center for Nuclear Physics and Department of Physics, Osaka University, Ibaraki, Osaka, Japan}
\address[unc]{Department of Physics and Astronomy, University of North Carolina, Chapel Hill, NC, USA}
\address[alberta]{Centre for Particle Physics, University of Alberta, Edmonton, AB, Canada}
\address[blhill]{Department of Physics, Black Hills State University, Spearfish, SD, USA}
\address[ttu]{Tennessee Tech University, Cookeville, TN, USA}
\address[sjtu]{Shanghai Jiao Tong University, Shanghai, China}
\fntext[TU]{Permanent address: Tuebingen University, Tuebingen, Germany}
\fntext[ucb]{Alternate Address: Department of Nuclear Engineering, University of California, Berkeley, CA, USA}

\begin{abstract}
 The \mj~Collaboration is constructing a system containing 40~kg of HPGe detectors to demonstrate the feasibility and potential of a future tonne-scale experiment capable of probing the neutrino mass scale in the inverted-hierarchy region. To realize this, a major goal of the~\mj~\dem~is to demonstrate a path forward to achieving a background rate at or below 1~cnt/(ROI-t-y) in the 4~keV region of interest around the Q-value at 2039~keV. This goal is pursued through a combination of a significant reduction of radioactive impurities in construction materials with analytical methods for background rejection, for example using powerful pulse shape analysis techniques profiting from the p-type point contact HPGe detectors technology. The effectiveness of these methods is assessed using simulations of the different background components whose purity levels are constrained from radioassay measurements.
\end{abstract}


\end{frontmatter}

\section{Motivation: neutrinoless double-beta decay}
\label{sec1}

Neutrinoless double-beta (0$\nu\beta\beta$) decay is a general, model independent method to search for lepton number violation and to determine the Dirac/Majorana nature of the neutrino~\cite{Zralek,Camilleri,Avignone}. Observation of this rare process would have significant implications for our understanding of the nature of neutrinos and matter in general. The 0$\nu\beta\beta$-decay rate may be written as:

\begin{equation}\label{eq1}
\left(T^{0\nu}_{1/2}\right)^{-1}=G^{0\nu}|M_{0\nu}|^{2} \left( \frac{\langle m_{\beta\beta}\rangle}{m_{e}} \right)^{2}
\end{equation}
\noindent
where $G^{0\nu}$ is a phase space factor including the couplings, $M_{0\nu}$ is a nuclear matrix element, $m_{e}$ is the electron mass,
and $\langle m_{\beta\beta}\rangle$ is the effective Majorana neutrino mass. The latter is given by

\begin{equation}\label{eq2}
\langle m_{\beta\beta}\rangle=\left|\sum_{i=0}^{3}U_{ei}^{2}m_{i}\right|
\end{equation}
\noindent
where $U_{ei}$ specifies the admixture of neutrino mass eigenstate $i$ in the electron neutrino. Then, assuming that 0$\nu\beta\beta$-decay is mainly driven by the exchange of light Majorana neutrinos, it is possible to establish an absolute scale for the neutrino mass, provided that the nuclear matrix elements are known.

Experimentally, 0$\nu\beta\beta$-decay can be detected by searching the spectrum of the summed energy of the emitted betas for a monoenergetic line at the Q-value of the decay (\QBB). In previous-generation searches, the most sensitive limits on 0$\nu\beta\beta$-decay came from the Heidelberg-Moscow experiment~\cite{Heil}, and the IGEX experiment~\cite{IGEX,IGEX2}, both using $^{76}$Ge. A direct observation of 0$\nu\beta\beta$-decay was claimed by a subgroup of the Heidelberg-Moscow collaboration~\cite{Kla}. Recent sensitive searches for~\BBz~have been carried out in $^{76}$Ge (GERDA~\cite{GERDA}) and $^{136}$Xe (KamLAND-Zen~\cite{KamLAND} and EXO-200~\cite{EXO,EXO2}), setting sensitive limits that do not support such a claim.

\section{The \textsc{ Majorana Demonstrator}}
\label{sec2}

The~\mj~\dem~\cite{mjd} is an array of enriched and natural germanium detectors that will search for the 0$\nu\beta\beta$-decay of $^{76}$Ge. The specific goals of the~\mj~\dem~are:

\begin{itemize}
  \item Demonstrate a path forward to achieving a background rate at or below 1~cnt/(ROI-t-y) in the 4~keV region of interest (ROI) around the 2039~keV~\QBB~of the $^{76}$Ge 0$\nu\beta\beta$-decay.
  \item Show technical and engineering scalability toward a tonne-scale instrument.
  \item Field an array that provides sufficient sensitivity to test the Klapdor-Kleingrothaus claim and to be comparable with alternate approaches.
  \item Perform searches for other physics beyond the standard model, such as dark matter and axions.
\end{itemize}

The~\mj~\dem~will be composed of 40~kg of HPGe detectors which also act as the source of $^{76}$Ge \BBz-decay. The benefits of HPGe detectors include intrinsically low-background source material, understood enrichment chemistry, excellent energy resolution, and sophisticated event reconstruction. P-type point contact (PPC) detectors that allow powerful background rejection were chosen after extensive R\&D by the collaboration. The baseline plan calls for 30~kg of the detectors to be built from Ge material enriched to 87\% in isotope 76 and 10~kg fabricated from $^{\rm nat}$Ge (7.8\% $^{76}$Ge). Each detector has a mass of about 0.6-1.0~kg. The main technical challenge is the reduction of environmental ionizing radiation backgrounds by about a factor of 100 below what has been achieved in previous experiments. A modular instrument composed of two cryostats built from ultra-pure electroformed copper has been designed. Each module will host 7 strings of detectors: each detector, with cylindrical shape, is housed in a frame referred to as a detector unit, and up to five detector units are stacked into a string. The modules will be operated in a graded passive shield, which is surrounded by a 4$\pi$ active muon veto. The modular approach will allow the assembly and commissioning of each module independently, providing a fast deployment with minimum interference on already-operational detectors.

At present, the~\mj~\dem~is under construction: the shielding is being installed, the HPGe detectors are being characterized, and the Prototype Module, an initial prototype cryostat fabricated from commercially produced copper, is taking data with two strings of detectors produced from natural germanium. It serves as a test bench for mechanical designs, fabrication methods, and assembly procedures that will be used for the construction of the electroformed-copper Modules~1~\&~2. The first module of enriched germanium detectors will be assembled in 2014. The second enriched Ge module will be subsequently implemented, completing the~\mj~\dem.~The beginning of the data taking with Module~1 is foreseen for 2015.

\section{Background Model for the \textsc{Majorana Demonstrator}}
\label{sec3}

The expected background in the~\mj~\dem,~4.1~cnts/(ROI-t-y) after analysis cuts, will project to a background level of 1~cnt/(ROI-t-y) in a large scale experiment after accounting for additional modest improvements from thicker shielding, better self-shielding, and if necessary, increased depth. This background level represents a substantial improvement over previous generation experiments. To achieve this goal, backgrounds must be both reduced at the source and rejected in the analysis.

\begin {figure}[ht]
\includegraphics[width=0.5\textwidth]{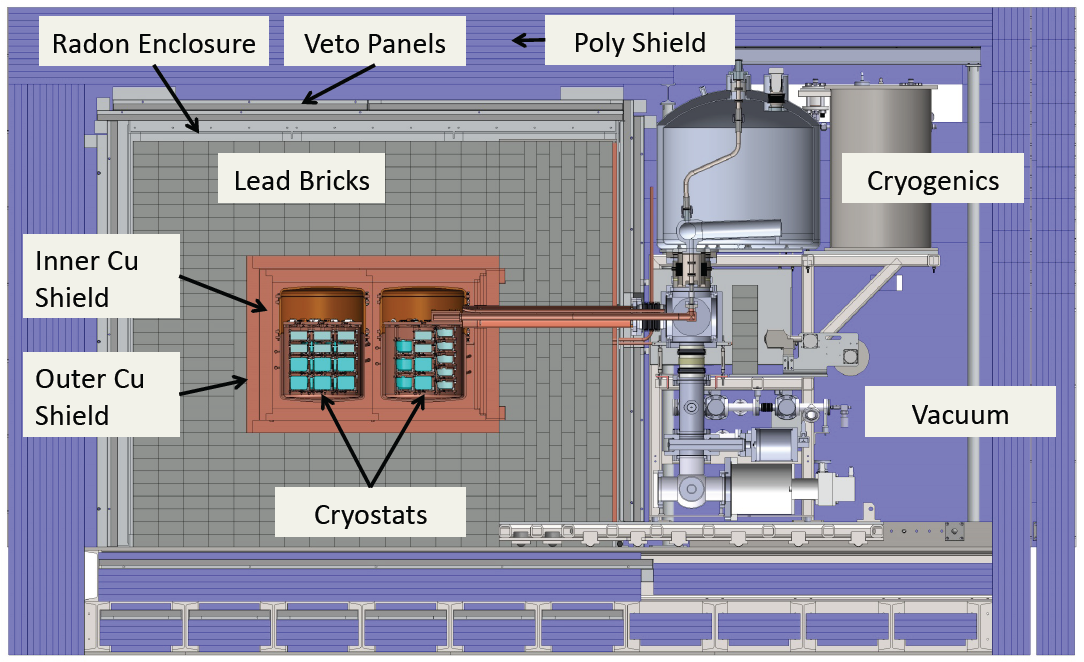}
\centering \caption{\it Schematic drawing of the \textsc{Majorana Demonstrator} shown with both modules installed. The different layers of the shield are indicated in the drawing. }
\label{fig:shie}
\end {figure}

 One significant means of background reduction is the fielding of the detectors in large arrays that share a cryostat, minimizing the amount of interstitial material. Reduction is also achieved by better shielding of the detector from environmental radioactivity, and by the reduction of radioactive impurities in construction materials, including minimization of exposure to cosmic rays. The low-radioactivity passive shield has an innermost layer of 5-cm-thick of electroformed copper, a subsequent layer of 5-cm-thick Oxygen-Free High thermal Conductivity (OFHC) copper, a 45-cm-thick high-purity lead shield, and 30-cm thickness of polyethylene with the inner layer borated to act as a neutron moderator. The shield is encapsulated by an anti-radon system, and an active veto system comprised of two layers of scintillator panels is used to detect muons. The entire apparatus is located and placed at the Davis Campus of the Sanford Underground Research Facility on the 4850' level (1478~m) in Lead, South Dakota, to avoid cosmic rays. Figure~\ref{fig:shie} shows a schematic drawing of the different layers of the shield.

The high material purities required for the \textsc{Majorana Demonstrator} necessitated the development of improved assay capabilities. These capabilities are needed not just to establish that the required purities can be achieved, but to also monitor construction processes to verify that cleanliness is maintained. The three primary assay methods used are: gamma-ray counting, inductively coupled plasma mass spectrometry (ICP-MS), and neutron activation analysis (NAA). In the following the main aspects of the apparatus materials are summarized.% and in Figure~\ref{fig:2} their estimated background contributions are shown.

\begin{itemize}
\item The production process for enriched germanium detectors (enrichment, zone reﬁning, and crystal growth) efficiently removes natural radioactive impurities from the bulk germanium. The cosmogenic activation isotopes $^{60}$Co and $^{68}$Ge are produced in the crystals while they are above ground, but have been reduced by minimizing the time to deployment underground and by the use of passive shielding during transport and storage.
\item For the main structural material in the innermost region of the apparatus, copper was chosen for its lack of naturally occurring radioactive isotopes and its excellent physical properties. By starting with the cleanest copper stock identified and then electroforming it underground to eliminate primordial radioactivity and cosmogenically-produced $^{60}$Co, at least an order-of-magnitude background reduction over commercial alternatives has been achieved. Electroformed copper will also be employed for the innermost passive shield. Commercial copper stock is clean enough for use as the next layer of shielding.
\item Modern lead is available with sufficient purity to be used as the bulk shielding material outside of the copper layers.
\item Several clean plastics are available for electrical and thermal insulation. For the detector supports we use a pure Polytetraﬂuoroethylene (PTFE), DuPont$^{\rm TM}$~Teflon$^{\circledR}$~NXT-85. Thin layers of low-radioactivity parylene are being used as a coating on copper threads to prevent galling, and for the cryostat seal. For the few weight-bearing plastic components requiring higher rigidity, pure stocks of PEEK$^{\circledR}$~(polyether ether ketone), produced by Victrex$^{\circledR}$,~and Vespel SP-1$^{\circledR}$,~produced by DuPont$^{\rm TM}$~have been obtained.
\item The front-end electronics are also designed to be low mass and ultra-low background because they must be located in the interior of the array adjacent to the detectors in order to maintain signal fidelity. The circuit board is fabricated by sputtering thin traces of pure gold and titanium on a silica wafer, upon which a bare FET is mounted using silver epoxy. A $\sim$G$\Omega$-level feedback resistance is provided by depositing intrinsically pure amorphous Ge directly on the wafer. Detector contact is made via an electroformed copper pin with beads of low-background tin at either end. An electroformed-copper spring provides the contact force.
\item Signal and high-voltage cables are extremely low-mass miniature coaxial cable. An agreement with the vendor was created to cleanly fabricate the final product using pure stock that we provided for the conductor, insulation, and shield. Cable connectors within the cryostat are made with Mill-Max pins housed in Vespel$^{\circledR}$~bodies fabricated in-house.
\end{itemize}

After background reduction, background rejection takes place. One key advantage of HPGe detectors is their inherently excellent energy resolution ($<$0.2\% at $Q_{\beta\beta}$) associated with the low threshold for electron-pair production, leading to a narrow ROI (4~keV). As 0$\nu\beta\beta$-decay produces a mono-energetic peak in the measured spectrum at 2039~keV, improving the resolution reduces the continuum backgrounds in the ROI, allowing for a better identification of lines in the spectrum, and minimizing the contribution from leakage of the irreducible 2$\nu\beta\beta$-decay spectrum into the ROI. The~\mj~\dem~will also make use of event signatures, such as detector coincidences and time correlations, to reject events not attributable to 0$\nu\beta\beta$-decay. For example, background signals from radioactive decay often include a $\beta$ and/or one or more $\gamma$ rays that can scatter into different detectors, and these decays sometimes occur in chains that result in time-correlated event signatures. PPC detectors have an additional ability to discriminate against such $\gamma$ rays, which frequently undergo multiple scatters over several centimeters within a Ge crystal. 0$\nu\beta\beta$-decay energy deposition, on the other hand, typically occurs within a small volume ($\sim$1~mm$^{3}$) giving a single site energy deposit.

The analysis of detector events is envisioned to proceed primarily as a series of cuts. First, the detector hit granularity allows to reject simultaneous events in coincidence in two or more crystals. Second, events in coincidence with a muon passing through the vetoes will be also discarded. Third, PPC detectors possess superb Pulse Shape Analysis (PSA) discrimination between single-site interactions and multi-site interaction events separated by more than a few mm within the crystal bulk, making them highly suitable for 0$\nu\beta\beta$ searches. Fourth, the low energy threshold of the detectors makes them suitable for event time-correlation discrimination using x-rays. For example, $^{68}$Ga decays can only result in background to 0$\nu\beta\beta$-decay if one of the annihilation $\gamma$ rays interacts in the same crystal that contains the $\beta^{+}$. Hence it is always a multiple-site energy deposit and we reject much of this background through PSA. In addition, however, $^{68}$Ga decays are preceded by the electron capture decay of the parent $^{68}$Ge. The low threshold of the PPC detectors permits additional rejection by a time-correlation cut with the $^{68}$Ge $\sim$10~keV K-shell and $\sim$1~keV L-shell x-rays.

To study the sensitivity of Ge detector array designs to various background sources, the~\mj~and GERDA collaborations have jointly developed a simulation software framework, MaGe~\cite{mage}, that is based on the Geant4~\cite{Geant4} simulation toolkit. MaGe is used to simulate the response of ultra-low radioactive background HPGe detectors to ionizing radiation. The MaGe framework contains the geometry models of common detector objects, experimental prototypes, test stands, and the~\mj~\dem~itself. In the prototyping phase, the simulation was used as a virtual test stand guiding detector design, allowing an estimate of the effectiveness of proposed background reduction techniques, and providing an estimate of the experimental sensitivity. The simulations include estimates of the rejection efficiencies from the analysis cuts. The PSA cut efficiencies in particular are estimated using a heuristic calculation in which multiple interactions in a detector are examined for their relative drift time using isochrone maps. That information is combined with the energies of the interactions to determine whether the PSA algorithm would be capable of rejecting the event.

\begin {figure}[ht]
\includegraphics[width=0.5\textwidth]{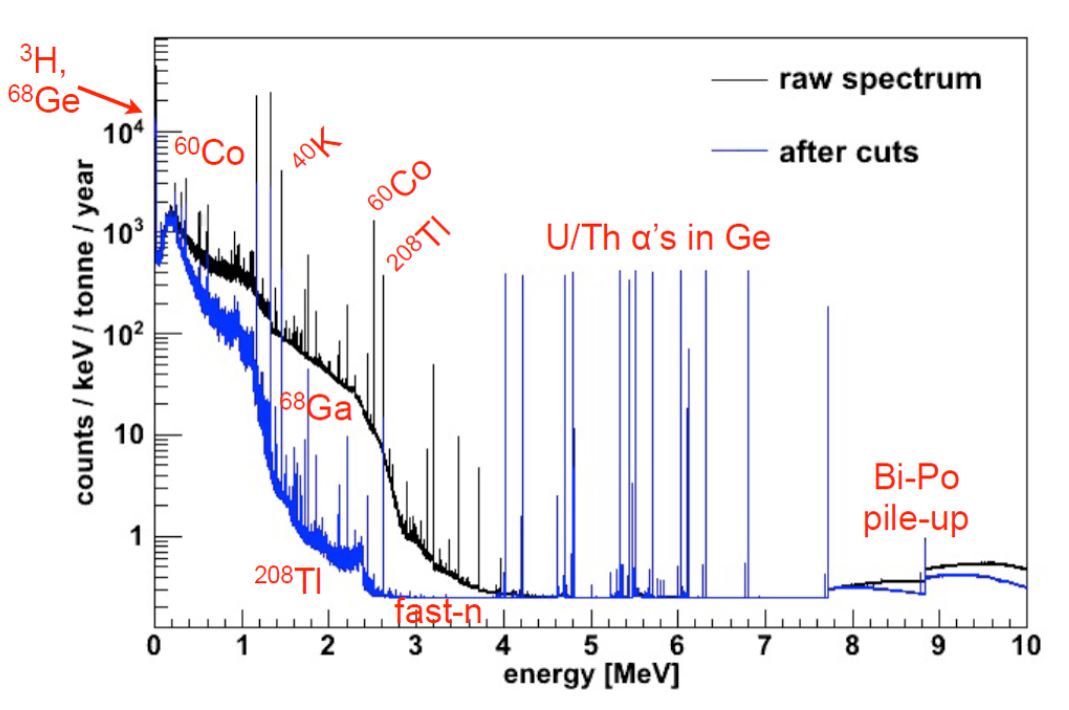}
\centering \caption{\it Simulated background spectrum of the~\mj~\dem~before (black) and after (blue) analysis cuts. The selected binning is 1~keV.}
\label{fig:bkg}
\end {figure}

\begin {figure}[ht]
\includegraphics[width=0.55\textwidth]{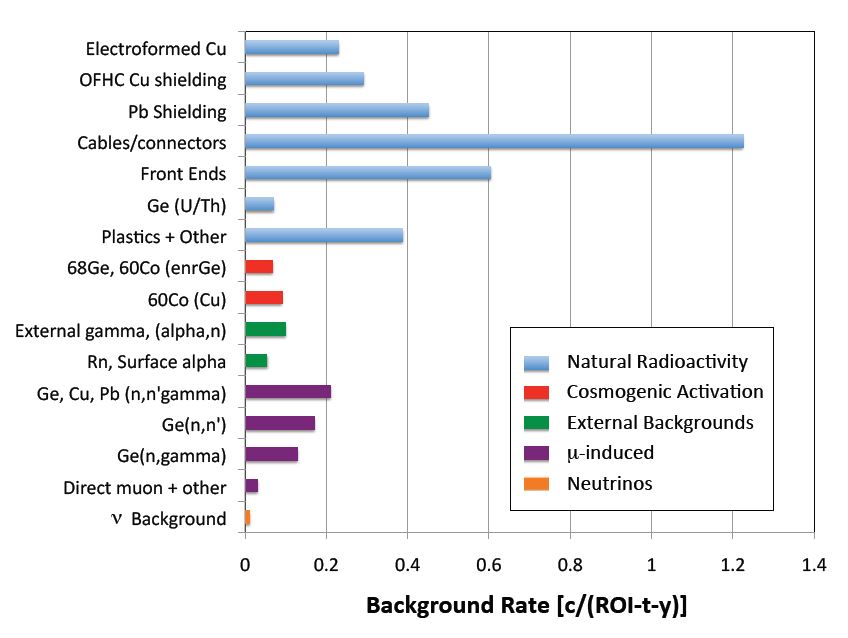}
\centering \caption{\it Estimated ROI background contributions for the \BBz-decay search shown in cnts/(ROI-t-y). Backgrounds from natural radioactivity from the detector materials are shown in blue. Cosmogenic activation products are colored red. Green denotes backgrounds from the environment or those introduced during detector assembly. Purple is for muon-induced backgrounds at depth. An upper limit on the negligible background from atmospheric and other neutrinos is shown in orange. The contributions sum to 4.1~cnts/(ROI-t-y) in the~\mj~\dem. This projects to a background level of $\sim$1~cnt/(ROI-t-y) in a large scale experiment after accounting for additional modest improvements from thicker shielding, better self-shielding, and if necessary, increased depth.}
\label{fig:2}
\end {figure}

The projected background spectrum before and after all cuts is given in Figure~\ref{fig:bkg}. A summary of the contributions to the 0$\nu\beta\beta$-decay ROI is given in Figure~\ref{fig:2}. These simulation results were used for the current background estimates and will be used in future analyses of data. For that purpose, a flexible, as-built simulated geometry has been developed. This geometry will be used in upcoming simulations of the Prototype Module and Module 1 to quantify the calibration requirements and analyze calibration data taken during commissioning. During operation, MaGe will be used to simulate and characterize backgrounds to determine the ultimate sensitivity of the experiment. It will also provide probability distribution functions (PDFs) for likelihood-based signal extraction analyses. The collaboration has also developed an automated validation suite that thoroughly tests all of the critical physics process that are being simulated by MaGe and Geant4 against experimental data. This suite is run every time there is a major update to MaGe or Geant4 to verify that the critical physics processes do not show degraded performance between versions.

\section{Sensitivity}
\label{sec4}

\begin {figure}[ht]
\includegraphics[width=0.55\textwidth]{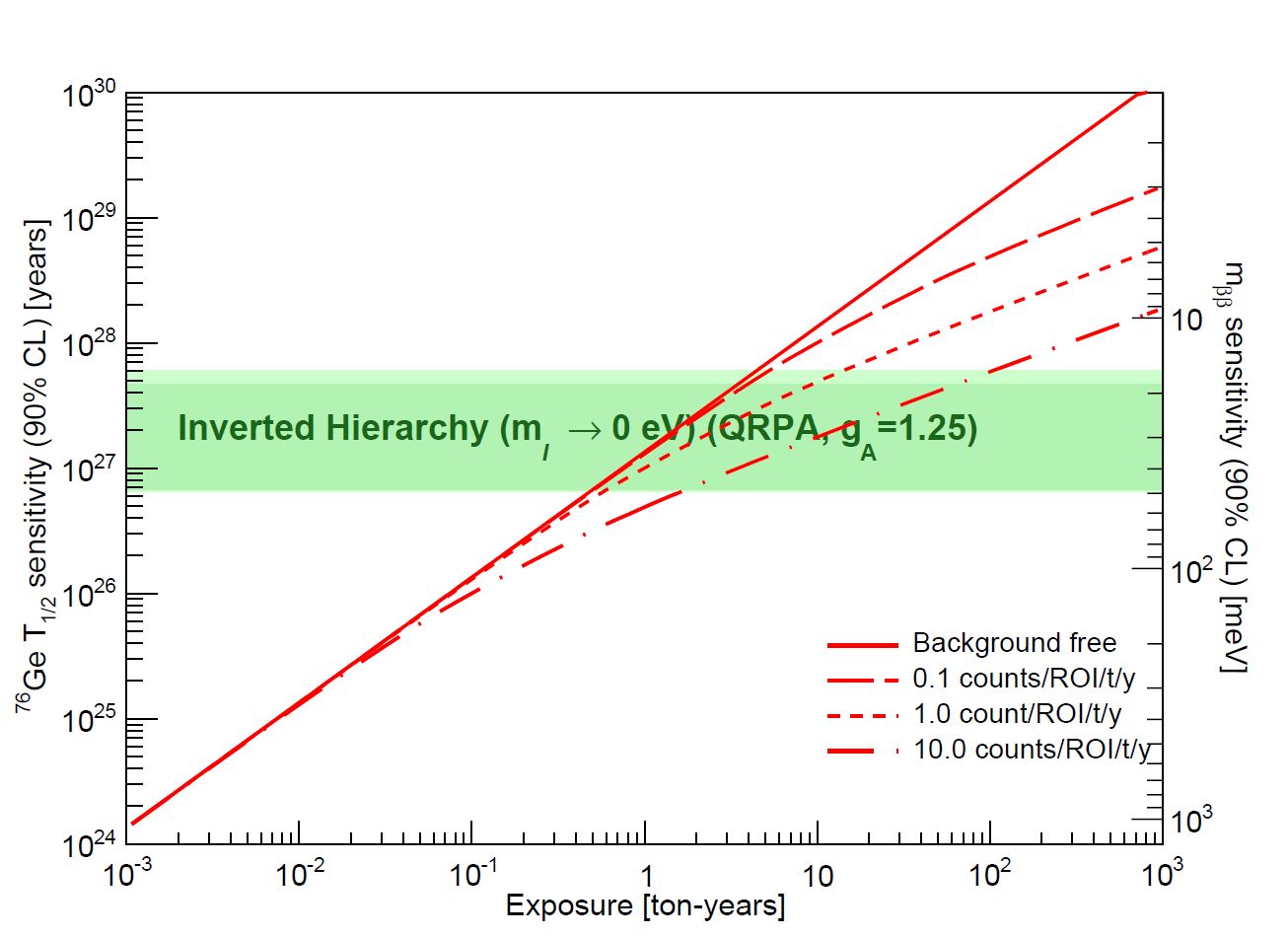}
\centering \caption{\it 90\% C.L. sensitivity as a function of exposure for 0$\nu\beta\beta$-decay searches in $^{76}$Ge under different background scenarios. The matrix element from Ref.~\cite{MatEl} was used to convert from half-life to neutrino mass. The upper shaded band shows the region where a signal would be detected should the Klapdor-Kleingrothaus claim~\cite{Kla} be correct. $m_{l}$ in the lower shaded band refers to the lightest neutrino mass.}
\label{fig:sens}
\end {figure}

The sensitivity of a 0$\nu\beta\beta$ search increases with the exposure of the experiment, but ultimately depends on the achieved background level. This relationship is illustrated in Figure~\ref{fig:sens} with $^{76}$Ge, where the Feldman-Cousins definition of sensitivity has been used to transition smoothly between the background-free and background-dominated regimes. In order to reach the neutrino mass scale associated with the inverted mass hierarchy, 15-50~meV, a half-life sensitivity greater than 10$^{27}$~y is required. This corresponds to a signal of a few counts or less per tonne-year in the 0$\nu\beta\beta$ peak. Observation of such a small signal will require tonne-scale detectors with background contributions at or below a rate of 1~cnt/(ROI-t-y). HPGe detectors have exceptionally good intrinsic energy resolution of $\sim$0.2\% at Q$_{\beta\beta}$ of 2039~keV, which for a 4~keV region of interest would correspond to a required background of $<$0.25~cnts/(keV-t-y). This excellent energy resolution ensures that background from the irreducible 2$\nu\beta\beta$-decay (T$_{1/2}$\,=\,1.8\,x\,10$^{21}$~y) is negligible even for 0$\nu\beta\beta$ half lives beyond 10$^{27}$~y. Although this figure is drawn using experimental parameters and theoretical nuclear matrix elements relevant for 0$\nu\beta\beta$ searches using $^{76}$Ge, the situation for other isotopes is not qualitatively different. It may be concluded that achieving sensitivity to the entire parameter space for inverted-hierarchical Majorana neutrinos would require, using optimistic values of matrix elements and g$_{A}$, about 5-10 tonne-years of exposure with a background rate of less than 1~cnt/(ROI-t-y). Higher background levels would require signiﬁcantly more mass to achieve the same sensitivity within a similar counting time. Finally, we would like to emphasize that a convincing discovery that neutrinos are Majorana particles and that lepton number is violated will require the observation of 0$\nu\beta\beta$ in multiple experiments using different 0$\nu\beta\beta$ isotopes.

\newpage

\section{Acknowledgments}

We acknowledge support from the Office of Nuclear Physics in the DOE Office of Science, the Particle Astrophysics Program of the National Science Foundation, and the Russian Foundation for Basic Research. We acknowledge the support of the Sanford Underground Research Facility administration and staff.

\label{}



%% References

\bibliographystyle{unsrt}
\bibliography{CCuesta_Background}

\providecommand{\url}[1]{#1}  \newcommand{\noop}[1]{}
\begin{thebibliography}{10}

\bibitem{Zralek}
M.~Zralek.
\newblock {On the Possibilities of Distinguishing Dirac from Majorana
  Neutrinos}.
\newblock {\em ACTA Phys. Pol. B}, 88:2225, 1997.

\bibitem{Camilleri}
{L. Camilleri, E. Lisi and J.F. Wilkerson}.
\newblock {Neutrino Masses and Mixings: Status and Prospects}.
\newblock {\em Ann. Rev. Nucl. Part. Sci.}, 58:343, 2008.

\bibitem{Avignone}
{F. T. III Avignone, S. R. Elliott and J. Engel}.
\newblock {Double beta decay, Majorana neutrinos, and neutrino mass}.
\newblock {\em Rev. mod. Phys.}, 80:481, 2008.

\bibitem{Heil}
L.~Baudis et~al.
\newblock {Limits on the Majorana Neutrino Mass in the 0.1\,eV Range}.
\newblock {\em Phys. Rev. Lett.}, 83:41, 1999.

\bibitem{IGEX}
C.~E.~Aalseth et~al.
\newblock {IGEX $^{76}$Ge neutrinoless double-beta decay experiment: Prospects
  for next generation experiments}.
\newblock {\em Phys. Rev. D}, 65:092007, 2002.

\bibitem{IGEX2}
C.~E.~Aalseth et~al.
\newblock {The IGEX experiment reexamined: A response to the critique of
  Klapdor-Kleingrothaus, Dietz, and Krivosheina}.
\newblock {\em Phys. Rev. D}, 70:078302, 2004.

\bibitem{Kla}
H.~V. Klapdor-Kleingrothaus and I.~V. Krivosheina.
\newblock {The evidence for the observation of 0nu beta beta decay: The
  identification of 0nu beta beta events from the full spectra}.
\newblock {\em Mod. Phys. Lett. A}, 21:1547, 2006.

\bibitem{GERDA}
M.~Agostini et~al.
\newblock {Results on Neutrinoless Double-$\beta$ Decay of $^{76}$Ge from Phase
  I of the GERDA Experiment}.
\newblock {\em Phys. Rev. Lett.}, 111:122503, 2013.

\bibitem{KamLAND}
A.~Gando et~al.
\newblock {Limit on Neutrinoless $\beta\beta$ Decay of $^{136}$Xe from the
  First Phase of KamLAND-Zen and Comparison with the Positive Claim in
  $^{76}$Ge}.
\newblock {\em Phys. Rev. Lett.}, 110:062502, 2013.

\bibitem{EXO}
J.~B.~Albert et~al.
\newblock {Improved measurement of the 2νββ half-life of $^{136}$Xe with the
  EXO-200 detector}.
\newblock {\em Phys. Rev. C}, 89:015502, 2014.

\bibitem{EXO2}
J.~B.~Albert et~al.
\newblock {Search for Majorana neutrinos with the first two years of EXO-200
  data}.
\newblock {\em arXiv}, 1402.6956, 2014.

\bibitem{mjd}
N.~Agrall et~al.
\newblock {The \textsc{Majorana Demonstrator} Neutrinoless Double-Beta Decay
  Experiment}.
\newblock {\em Adv. High En. Phys.}, 2014:365432, 2014.

\bibitem{Geant4}
S.~Agostinelli et~al.
\newblock {GEANT4} a simulation toolkit.
\newblock {\em Nucl. Ins. Meth. A}, 506:250--303, 2003.

\bibitem{mage}
M.~Boswell et~al.
\newblock {MAGE-a GEANT4-Based Monte Carlo Application Framework for
  Low-Background Germanium Experiments}.
\newblock {\em IEEE Trans. Nucl. Sci.}, 58:1212, 2011.

\bibitem{MatEl}
F.~\v{S}imkovic~et al.
\newblock {0$\nu\beta\beta$-decay nuclear matrix elements with self-consistent
  short-range correlations}.
\newblock {\em Phys. Rev. C}, 79:055501, 2009.

\end{thebibliography}

\end{document}